\documentclass[twocolumn,english,aps,preprint,prl,twocolumn,reprint]{revtex4-1}
\usepackage[T1]{fontenc}
\usepackage[latin9]{inputenc}
\usepackage{textcomp}
\usepackage{bm}
\usepackage{amstext}
\usepackage{graphicx}
\usepackage{esint}

\makeatletter

\@ifundefined{textcolor}{}
{%
\definecolor{BLACK}{gray}{0}
\definecolor{WHITE}{gray}{1}
\definecolor{RED}{rgb}{1,0,0}
\definecolor{GREEN}{rgb}{0,1,0}
\definecolor{BLUE}{rgb}{0,0,1}
\definecolor{CYAN}{cmyk}{1,0,0,0}
\definecolor{MAGENTA}{cmyk}{0,1,0,0}
\definecolor{YELLOW}{cmyk}{0,0,1,0}
}

\makeatother

\usepackage{babel}
\begin{document}

\title{The scaling of boson sampling experiments}

\author{P. D. Drummond, B. Opanchuk, L. Rosales-Z\'arate, M. D. Reid }

\affiliation{Centre for Quantum and Optical Science, Swinburne University of Technology,
Hawthorn, Victoria 3122, Australia}

\author{P. J. Forrester}

\affiliation{Department of Mathematics and Statistics, ARC Centre of Excellence
for Mathematical \& Statistical Frontiers, The University of Melbourne,
Victoria 3010, Australia.}
\begin{abstract}
Boson sampling is the problem of generating a quantum bit stream whose
average is the permanent of a $n\times n$ matrix. The bitstream is
created as the output of a prototype quantum computing device with
$n$ input photons. It is a fundamental challenge to verify boson
sampling, and the question of how output count rates scale with matrix
size $n$ is crucial. Here we apply results from random matrix theory
to establish scaling laws for average count rates in boson sampling
experiments with arbitrary inputs and losses. The results show that,
even with losses included, verification of nonclassical behaviour
at large $n$ values is indeed possible.
\end{abstract}
\maketitle
Much recent attention has been given to the application of multichannel
linear photonic networks to solving computational tasks thought to
be inaccessible to any classical computer. Such devices are the prototypes
of quantum computers~\cite{KnillLaflammeMilburn,Kok2007} and novel
metrology devices~\cite{OBrienMetrology,Motes2015_PRL114}. Exponentially
hard problems that are not soluble with digital classical technology
have many potential applications~\cite{Papadimitriou:2003}. In particular,
``BosonSampling''~\cite{AaronsonArkhipov:2011,AaronsonArkhipov2013LV}
is the hard problem of how to generate a bitstream of photon-counts
with the distribution of a unitary or Gaussian permanent. This result
is created as the output from a single photon input to each of $n$
distinct channels. This is conjectured to be exponentially hard at
large $n$, while being relatively straightforward to implement physically.

It is widely appreciated that to \emph{verify} the solution is correct
is an important and significant challenge~\cite{Aaronson:2014,Carolan2014,Spagnolo2014,Gogolin2013,Bentivegna2015,AolitaEisert2015,Tichy2014,Bentivegna2016BSworking}.
The task is to measure the coincidence rates of counting $n$ photons
in $n$ output channels and to confirm that they correspond to the
modulus squared of an $n\times n$ sub-permanent of a unitary matrix.
The matrices are under experimental control~\cite{Broome2013,Crespi2013,Tillmann2013,Spring72013},
and an average over a random ensemble of them is necessary to verify
the experiments. A number of known strategies exist.

Yet, since permanents are intrinsically exponentially hard to compute~\cite{Valiant1979},
it is nontrivial to verify the output is correct~\cite{AaronsonArkhipov2013LV,Aaronson:2014}
at large $n$ values. This computational issue makes it difficult
to estimate how count rates scale unless averaged over all unitaries.
Understanding scaling is essential because count rates decline exponentially
fast as $n$ increases, requiring a strategy that overcomes this.

In this Letter, we solve the average scaling problem for arbitrary
inputs and losses. We obtain the maximum scaling of improvements that
are possible with recent channel grouping strategies~\cite{Carolan2014}.
To achieve this, we combine the generalized P-representation with
methods from random matrix theory to obtain averages over unitary
transformations. This allows to describe realistic photonic network
experiments, with arbitrary inputs, outputs, and losses. Such losses
in boson sampling have recently been investigated elsewhere \cite{Motes2015Losses,Aaronson2016_LostPhotons}.

The scaling improvement with channel grouping depends on the channel
occupation ratio, $k=m/n$, reaching over a hundred orders of magnitude
at $k=6$, $n=100$. This is well beyond the capability of any classical,
exact computation of matrix permanents. Our results show that boson-sampling
verification with such large $n$ values is possible provided high
efficiency detectors are available.

Scaling issues like this arising in random matrix theory are widespread~\cite{forrester2010log},
as averages over unitaries are fundamental to quantum physics. We
note an unexpected analogy with the statistics of a classical device
for generating random counts. On averaging over all unitaries, the
probability of $n$ single-photon counts in $n$ preselected channels~\textemdash{}
a \emph{quantum} Galton's board~\cite{Gard2014}~\textemdash{} is
identical to a classical Galton's board. The only difference is that
there are now $n-1$ additional virtual channels, which describe multi-photon
events in an output mode.

These extra channels can be thought of as non-classical communication
channels. Such channel capacity improvements are known from quantum
communication theory~\cite{caves1994quantum}, and are closely related
to Arkhipov and Kuperberg's ``birthday paradox'' for bosons~\cite{arkhipov2012bosonic}.

We start with a result~\cite{Glauber1963_CoherentStates,Mandel1995_book}
from quantum optics: any bosonic correlation function is obtainable
from the normally-ordered quantum characteristic function,
\begin{equation}
\chi\left(\bm{\xi}\right)=\left\langle :e^{\bm{\xi}\cdot\hat{\bm{a}}^{\dagger}-\bm{\xi}^{*}\cdot\hat{\bm{a}}}:\right\rangle .\label{eq:quant_char}
\end{equation}
Here, $\left\langle \hat{O}\right\rangle \equiv Tr\left[\hat{\rho}\hat{O}\right]$
is a quantum average, which we calculate using a generalized P-representation~\cite{Drummond_Gardiner_PositivePRep}.
This approach extends the Glauber P-function~\cite{Glauber_1963_P-Rep},
giving a distribution $P\left(\bm{\alpha},\bm{\beta}\right)$ over
are two $m$-component complex vectors, which exists for any $m-$mode
bosonic state $\hat{\rho}$. The quantum characteristic is then obtained
from \cite{Gardiner_Book_QNoise}
\begin{equation}
\chi\left(\bm{\xi}\right)=\int P\left(\bm{\alpha},\bm{\beta}\right)\chi\left(\bm{\xi}|\bm{\alpha},\bm{\beta}\right)d\mu\left(\bm{\alpha},\bm{\beta}\right)\,,\label{eq:P-functionchar}
\end{equation}
where $d\mu\left(\bm{\alpha},\bm{\beta}\right)$ is the integration
measure, and $\chi\left(\bm{\xi}|\bm{\alpha},\bm{\beta}\right)\equiv\exp\left(\bm{\xi}\cdot\bm{\beta}-\bm{\xi}^{*}\cdot\bm{\alpha}\right)$
is the conditional characteristic function for $\bm{\xi}$ given a
particular quantum phase-space trajectory $\bm{\alpha},\bm{\beta}$.

Transmission through a linear network changes the input density matrix
$\hat{\rho}^{(\mathrm{in})}$ to an output density matrix $\hat{\rho}^{(\mathrm{out})}$.
An amplitude transmission matrix $T$ transforms the coherent amplitudes~\cite{drummond2014quantum},
so that $\bm{\alpha}^{(\mathrm{out})},\,\bm{\beta}^{(\mathrm{out})}=T\bm{\alpha},\,T^{*}\bm{\beta}$.
The \emph{output} characteristic $\chi^{(\mathrm{out})}$ now depends
on the \emph{input} phase-space amplitude $\bm{\alpha},\bm{\beta}$
in an intuitively understandable way:
\begin{equation}
\chi^{(\mathrm{out})}\left(\bm{\xi}|\bm{\alpha},\bm{\beta}\right)=e^{\bm{\xi}\cdot\bm{T}^{*}\bm{\beta}-\bm{\xi}^{*}\cdot\bm{T}\bm{\alpha}}\,.\label{eq:unitary-characteristic}
\end{equation}

To calculate the average scaling behavior, we consider the case of
$T=\sqrt{t}U$, in which the unitary mode transformation $U$ of the
photonic network is combined with an absorptive transmission coefficient
$t$, representing losses and detector inefficiencies. We compute
the average output correlations over all possible unitaries, indicated
by $\left\langle \right\rangle _{U}$, from random matrix theory~\cite{fyodorov2006permanental}.
This allows one to evaluate averages of exponentials of the unitary
matrices in the conditional characteristic function, $\chi^{(out)}$
of Eq.~(\ref{eq:unitary-characteristic}).

The result is an averaged conditional characteristic
\begin{equation}
\left\langle \chi^{(\mathrm{out})}\left(\bm{\xi}|\bm{\alpha},\bm{\beta}\right)\right\rangle _{U}=\left(m-1\right)!\sum_{j=0}^{\infty}\frac{\left[-t\left|\bm{\xi}\right|^{2}\bm{\beta}\cdot\bm{\alpha}\right]^{j}}{j!\left(m-1+j\right)!}\,.
\end{equation}

Inserting this unitary average in Eq.~(\ref{eq:P-functionchar})
gives an exact solution for any averaged observable in the photonic
network, with arbitrary inputs and losses. The output photon statistics
depend on $\bm{\beta}\cdot\bm{\alpha}$, which is the phase-space
equivalent of the total \emph{input} photon number $\hat{N}$. As
a result, the characteristic function after unitary and quantum averaging
is:
\begin{equation}
\left\langle \chi^{(\mathrm{out})}\left(\bm{\xi}\right)\right\rangle _{U}=\left(m-1\right)!\sum_{j=0}^{\infty}\frac{\left(-t\left|\bm{\xi}\right|^{2}\right)^{j}\left\langle :\hat{N}^{j}:\right\rangle }{j!\left(m-1+j\right)!}\,.\label{eq:Exactchar}
\end{equation}

The effect of unitary averaging is such that all output channel and
phase information is lost, since the characteristic function now only
depends on $\left|\bm{\xi}\right|^{2}$. This also shows that all
output averages are obtained solely from the normally ordered input
photon number moments, $\left\langle :\hat{N}^{j}:\right\rangle $,
regardless of which input channels are used. As a common example,
for an input $n$-photon number state $\left\langle :\hat{N}^{j}:\right\rangle =n!/(n-j)!$,
so the sum in Eq.~(\ref{eq:Exactchar}) vanishes for $j>n$. This
is a consequence of photon-number conservation and the purely absorptive
loss reservoirs.

Only the photon number observables are non-vanishing after unitary
phase-averaging. These are most readily obtained from taking derivatives
of the photon-number generating function~\cite{cantrell1971generating},
\begin{equation}
G\left(\bm{\gamma}\right)\equiv Tr\left(\hat{\rho}\prod_{i}\left(1-\gamma_{i}\right)^{\hat{n}_{i}}\right)\,.
\end{equation}
Using the relationship between photon-number generator and characteristic
function~\cite{rockower1988calculating}, we find that:

\begin{equation}
G\left(\bm{\gamma}\right)\equiv\left(m-1\right)!\sum_{j=0}^{\infty}\frac{\left(-t\right)^{j}\left\langle :\hat{N}^{j}:\right\rangle }{\left(m-1+j\right)!}\sum_{\sum\mathbf{j}=j}\gamma_{1}^{j_{i}}\ldots\gamma_{m}^{j_{m}}
\end{equation}

This result is completely general for a photonic network with arbitrary
inputs, outputs and losses. A case of special interest is $P_{n|m}$,
the probability of observing $1$ photon in each of $n$ channels,
given an $n$-photon input and an $m$-mode network. This is found
on taking $n$ first derivatives of $G\left(\bm{\gamma}\right)$,
so that:
\begin{equation}
P_{n|m}=\frac{t^{n}\left(m-1\right)!n!}{\left(m-1+n\right)!}=t^{n}\left[C_{n}^{m+n-1}\right]^{-1}\,\label{eq:Coinicidence probability}
\end{equation}

We now wish to relate these results to the permanent of the transmission
matrix $T$. The permanent is a sum over all permutations $\sigma$
of the matrix indices of the product of $n$ terms, in which neither
row nor column indices are repeated. It is an exponentially hard object
to compute, and is one of the fundamental quantities addressed in
boson sampling theory and in linear optical networks \cite{Scheel2005,Scheel2004Permanents}.

For a pure, unitary state evolution, the photon counting probability
is the permanent of a sub-matrix, $\left\langle \left|\mathrm{perm}(U_{n|m})\right|^{2}\right\rangle $,
where $U_{n|m}$ is any $n\times n$ sub-matrix of $U$~\cite{Aaronson2011}.
More generally, we replace $U_{n|m}\rightarrow T_{n|m}$, so as to
include losses.

The permanent of a sub-matrix of $T$ is obtainable~\cite{li2015permanental}
from the permanental polynomial, which has similarities with moment
generating function. This is given by:

\begin{equation}
p(x)=\mathrm{perm}(xI-T)\equiv\sum_{n=0}^{m}b_{n}x^{m-n}\,.\label{eq:permanentalpoly}
\end{equation}

Next, we consider how to compute the unitary average of products of
the permanents, using$\left\langle .\right\rangle _{U}$ to indicate
averages over the circular unitary ensemble, with a Haar measure.
This is achieved through an elegant result in random matrix theory~\cite{fyodorov2006permanental}.
The unitary average of permanental polynomials in Eq.~(\ref{eq:permanentalpoly})
is:
\begin{eqnarray}
\left\langle p(x)p(y)^{*}\right\rangle _{U} & = & m!(m\text{\textminus}1)!\sum_{j=0}^{m}\frac{t^{j}\left(xy^{*}\right)^{m-j}}{\left(m-j\right)!(m-1+j)!}\,.
\end{eqnarray}

If $\omega^{(n)}=(\omega_{1},\ldots\omega_{n})$ where $\omega_{1}<\omega_{2}\ldots<\omega_{n}$,
we can define an $n\times n$ sub-matrix $T_{\bm{\omega}}\equiv T_{\omega_{i}\omega_{j}}$.
The coefficients of this polynomial are simply the sums over the permanents
of all possible distinct sub-matrices:
\begin{equation}
b_{n}=(-1)^{n}\sum_{\bm{\omega}^{(n)}}\mathrm{perm}(T_{\bm{\omega}^{(n)}})\,.
\end{equation}

The number of sub-matrices in the sum has a multiplicity given by
the binomial coefficient $C_{n}^{m}=m!/\left(n!(m-n)!\right)$, corresponding
to the different ways to choose the distinct indices $\omega_{j}$.
These indices have a straightforward physical interpretation: they
are the channel numbers of the input or output modes of the photonic
device.

Expanding the product average, and noting that products of different
sub-matrices vanish under unitary ensemble averaging, we consider
the sub-permanents with $j=n$. The average sum over all possible
sub-permanents of this size is a ratio of two binomial coefficients,
which we define as $R_{n|m}$:
\begin{equation}
\sum_{\bm{\omega}^{(n)}}\left\langle \left|\mathrm{perm}(T_{\bm{\omega}^{(n)}})\right|^{2}\right\rangle _{U}=R_{n|m}=\frac{t^{n}m!(m\text{\textminus}1)!}{(m-n)!(m+n-1)!}\,.\label{eq:sum-of-subs}
\end{equation}

All the averages are the same for every sub-matrix. Therefore, we
can replace $T_{\bm{\omega}^{(n)}}$ by any particular sub-matrix
$T_{n|m}$, and make use of the sub-matrix multiplicity. The final
result is the same as in Eq.~(\ref{eq:Coinicidence probability}),
except expressed using permanents, so that $P_{n|m}=\left\langle \left|\mathrm{perm}(T_{n|m})\right|^{2}\right\rangle _{U}$.
In the lossless limit of $t=1$, this result agrees with the ``bosonic
birthday paradox'' of Arkhipov and Kuperberg~\cite{arkhipov2012bosonic},
derived using different techniques.

We turn next to some limiting cases for large $n$, where $\log P_{n|m}\approx n\epsilon$
for a scaling exponent $\epsilon$.

\paragraph{Entire matrix}

If the matrix is the entire transmission matrix, then $n=m$. The
scaling exponent is $\epsilon=\log\left(t/4\right)$, and
\begin{equation}
\log P_{n|m}\,\mathop{\sim}\limits _{n\to\infty}\,\,n\epsilon+\frac{1}{2}\log\left[4\pi n\right]\,.
\end{equation}
This result generalizes one of Fyodorov~\cite{fyodorov2006permanental}.

\paragraph{Gaussian limit }

Next, take $n\ll m$, so that $k\gg1$. Standard methods for approximating
a binomial coefficient in this limit give the scaling exponent $\epsilon=\log\left(t/\left(k+1/2\right)\right)-1$,
where $k=m/n$, so that:
\begin{equation}
\log P_{n|m}\,\,\mathop{\sim}\limits _{n\to\infty}\,\,n\epsilon+\frac{1}{2}\log\left[2\pi n\right]\,.\label{eq:Gaussian}
\end{equation}
This is consistent with the fact that for large $k$, unitary sub-matrices
reduce to matrices with complex Gaussian random entries~\cite{jiang2009approximation,AaronsonArkhipov2013LV}.

\paragraph*{General sub-matrix}

For the general case, the scaling exponent is $\epsilon=\log t+k\log k-(1+k)\log(1+k)$,
and the asymptotic result is:

\begin{equation}
\log P_{n|m}\,\mathop{\sim}\limits _{n\to\infty}\,\,n\epsilon+\frac{1}{2}\log\left[2\pi n(1+1/k)\right]
\end{equation}

\begin{figure}
\includegraphics[width=1\columnwidth]{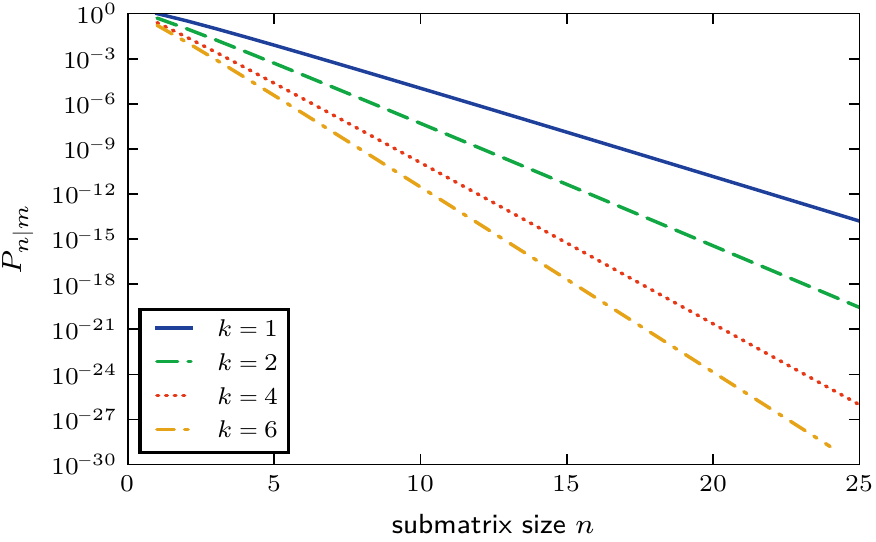}\protect\caption{Average sub-unitary permanent squared $P_{n|m}$ with $t=1$ for $k=1,2,4,6$,
with $k=1$ at the top and $k=6$ at the bottom.\label{fig:Average-subunitary-permanent}}
\end{figure}

The exact result is plotted in Figure~\ref{fig:Average-subunitary-permanent},
with different values of $k=m/n$. The power law is so close that
it cannot be told apart from the exact result on this scale. For all
numerical results, we choose $t=1$, since results in more realistic
cases with losses are readily obtained by adding $\log t$ to the
scaling exponents.

At large $k$ values, one obtains the Gaussian limit of Eq.~(\ref{eq:Gaussian}).
This gives an increasingly negative exponent, with exponentially small
count-rates.

Next, we show the result of an actual average over a finite, random
ensemble with $S=40000$ unitary samples. In the graphs, for $k=m/n=2$
and $n\leq25$ we give the relative errorWes in the asymptotic approximation
and a numerical average, compared to the exact solution. To estimate
statistical error-bars, we take an ensemble $S$, and divide it into
$\sqrt{S}$ sub-ensembles, giving sub-ensemble means which are approximately
Gaussian from the central limit theorem.

These are averaged, and the error in the mean $\sigma_{m}$ is obtained
using standard techniques. The error-bars are given in the plots as
$\pm\sigma_{m}$. The results agree with the exact equation with a
relative error comparable to the sampling error-bars. For the plotted
ratio of $k=2$, the errors are around $\pm1\%$ for over twenty orders
of magnitude range of values, and we see that the relative sampling
error over unitaries is independent of matrix size for $n\ge10$.

\begin{figure}
\includegraphics[width=1\columnwidth]{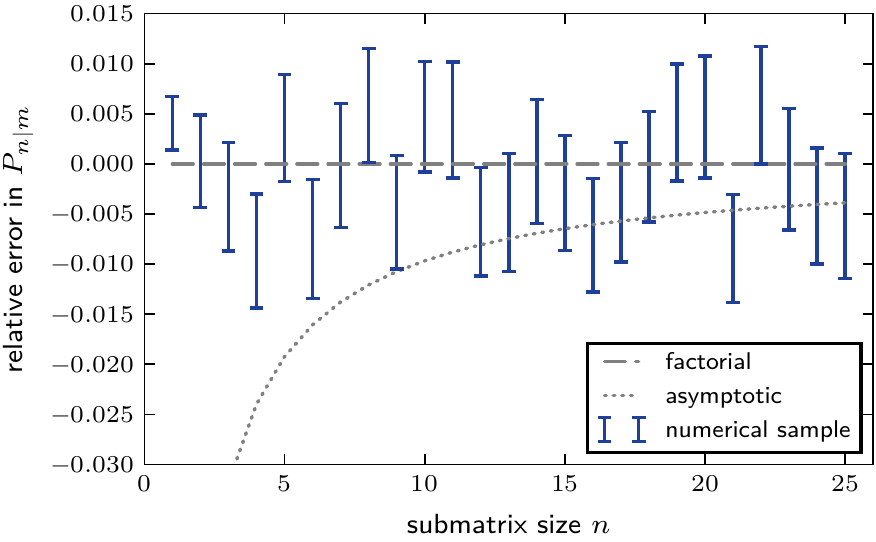}\protect\caption{Relative sampling errors in the sub-unitary permanent squared $P_{n|m}$
with $t=1$ for $k=2$. Numerical results for $40,000$ random unitaries
(solid lines with error bars) are compared to exact results (dashed
lines) and the asymptotic power law form (dotted line) for up to $n=25$.}
\end{figure}

While the scaling is better than the Gaussian limit of $k\gg1$, it
is still a problem for boson sampling verification. Even in the unlikely
case of perfect efficiency, the average permanent for a photon number
of $n=24$ is $\sim10^{-18}$ with a sub-matrix ratio of $k=2$. This
is the probability of a coincidence count, so one needs $10^{18}$
samples to obtain one count for a typical unitary. At a repetition
rate of $10^{12}\,\mathrm{Hz}$, which is the maximum one can reasonably
expect from the technology, one would require $10^{6}\,\mathrm{s}$
of measurement time for each count.

The reason for this is simple: many-body complexity. There are too
many quantum states possible. Monitoring the coincidence channels
for one many-body state takes too long, even though it is these counts
that are of interest. This property, although making verification
hard, is the most interesting feature of these experiments. They give
a uniquely controllable access to a laboratory system in which one
can unravel the complexity of a many-body system, in order to examine
each state, or arbitrary combinations of the quantum states.

Accordingly, suppose we consider what happens when one groups multiple
output channels together, by using logic gate operations on the detector
circuits, as in a recent, pioneering experiment~\cite{Carolan2014}.
A large number of randomized sets of channels can be combined, to
obtain a unique distinguishing signature for each unitary.

This strategy has important advantages over previous proposals. It
increases count-rates by exponentially large factors, and may allow
a test of the unitary output bitstream for matrices with permanents
larger than $n=40$, beyond the classical computation limits. Yet,
it is not restricted to any particular unitary, reducing the chance
that the device may only work in special cases. One can also use the
strategy to efficiently test for null counts, in edge cases like Fourier
matrices~\cite{Tichy2014}.

Our scaling laws predict the upper bound of the count-rate gain that
can achieved through sub-matrix multiplicity. The upper bound from
channel grouping is given by $R_{n|m}$, in Eq.~(\ref{eq:sum-of-subs}).
This has an exponent of $\lambda=\log t+2k\log k-(k-1)\log(k-1)-(k+1)\log(k+1)$,
so that the scaling is:

\begin{equation}
\log R_{n|m}\,\,\mathop{\sim}\limits _{n\to\infty}\,\,n\lambda+\frac{1}{2}\log\left[\frac{k+1}{k-1}\right]
\end{equation}

\begin{figure}
\includegraphics[width=1\columnwidth]{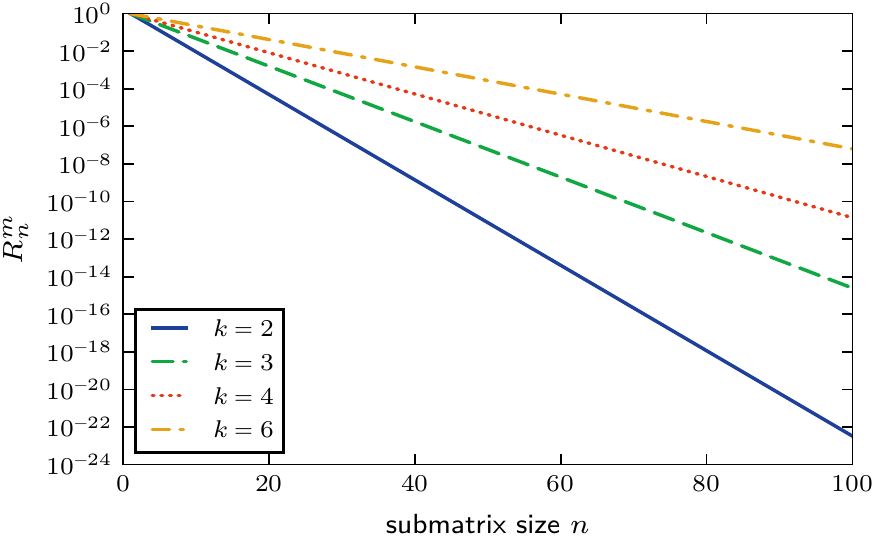}\protect\caption{Upper bound on count-rates $R_{n}^{m}$ for $n$ photons occurring,
without bunching, in any $n$ output channels, for $k=2,\ldots6$,
with $k=6$ at the top.\label{fig:Average-subunitary-bound}}
\end{figure}

For the Gaussian limit of $n,k\gg1$, one finds that $\lambda\rightarrow\log t-1/k$.
Unlike the single coincidence case, the grouped channel count rate
is \emph{maximized} for large $k$, rather than minimized as before.
The corresponding upper-bound result is plotted in Figure~\ref{fig:Average-subunitary-bound},
for different values of $k=m/n$, again taking $t=1$ for simplicity.

At at $k=6$, and $n=100$, there is now a dramatic increase of more
than $100$ orders of magnitude in total count-rate. The improvement
is greatest for large $k$ values, which are the cases of most interest.
High $n$ verification still requires total efficiencies of above
$90\%$, which is possible as the technology improves.

It is an intriguing result, mathematically and physically, that the
quantum Galton's board has a close relationship with the binomial
coefficients normally found in classical combinatorics. How can we
interpret this result?

Suppose that we replaced the photonic network, equivalent to a unitary
transformation, by an updated Galton's board device, which simply
switched the photons from the $n$ input channels to the $n$ output
channels in a random way. This would not involve interference, and
would have a similar behavior to a mechanical board, apart from an
increased number of inputs.

Under these conditions, the average probability of all counts occurring
in a preselected set of $n$ output channels is an inverse binomial
$\left[C_{n}^{m}\right]^{-1}$. We now see a truly remarkable result.
Apart from losses, the quantum Galton's board has, on averaging over
all unitaries, identical output coincidence probabilities to a classical
Galton's board with a number of channels given by $\tilde{m}=m+n-1$.

In other words, the fact that photons can bunch~\textemdash{} one
channel may carry up to $n$ photons, after all~\textemdash{} has
a similar effect on the output statistics as if the device were classical,
but with $n-1$ additional channels available. Needless to say, these
virtual channels do not exist. They represent, on average, the additional
output possibilities available owing to the fact that several bosons
can occupy the same mode, and hence occur in the same channel. This
is the large-scale consequence of the famous Hong-Ou-Mandel effect
in quantum optics~\cite{hong1987measurement,walborn2003multimode}.

It is this additional, virtual channel capacity that allows more quantum
information to be transmitted in a quantum photonic network than is
feasible if each channel was used separately, with one bit per channel.
Such extra capacity is a fundamental and important property of quantum
photonic networks~\cite{caves1994quantum}.

In summary, the unitary average of moduli of sub-permanents has remarkable
properties. Each permanent itself has no simple closed-form expression,
and one might imagine that taking an average over all possible unitaries
would only make things harder. Yet the average over the unitary ensemble
is just as simple as the closed form expression applicable to a classical
Galton's board.

As well as improved measurements, verification for an individual large
unitary requires improved computational methods, as standard methods
take exponentially long times. These results will be treated elsewhere.
In addition to understanding the scaling laws for verifying boson
sampling, our results may suggest how random matrix theory can be
applied to other, large-scale quantum technologies.
\begin{acknowledgments}
We wish to thank the Australian Research Council for their funding
support.
\end{acknowledgments}

\bibliography{Refs}

\end{document}